# Entropy cost of "Erasure" in Physically Irreversible Processes

R. E. Kastner * and A. Schlatter

* Correspondence: rkastner@umd.edu; November 11, 2023

**Abstract:** A restricted form of Landauer's Principle, independent of computational considerations, is shown to hold for thermal systems by reference to the joint entropy associated with conjugate observables. It is shown that the source of the compensating entropy for irreversible physical processes is due to the ontological uncertainty attending values of such mutually incompatible observables, rather than due to epistemic uncertainty as traditionally assumed in the information-theoretic approach. In particular, it is explicitly shown that erasure of logical (epistemic) information via reset operations is not equivalent to erasure of thermodynamic entropy, so that the traditional, information-theoretic form of Landauer's Principle is not supported by the physics. A further implication of the analysis is that, in principle, there can be no Maxwell's Demon in the real world.

## 1. Introduction.

Landauer's Principle (LP) was original presented by Landauer in terms of computation. Specifically, Landauer (1961) proposed that "computing machines" engaging in logically irreversible steps incur a cost of the order of kT per each step. While LP has been largely accepted, there is a dissenting minority (e.g., Earman and Norton 1999, henceforth 'EN'; Norton 2005-2018). While the present authors join the dissenters in disputing the identification of physical irreversibility with logical/computational irreversibility inherent in Landauer's original proposal, we nevertheless adduce a physical basis for a restricted form of LP: one which is not identified with computation, but with a narrower class of genuinely irreversible physical processes. One might consider this a form of Szilard's Principle if measurement is a physically irreversible process; the present work suggests that it is.

In the tradition of Shannon's work (Shannon, 1948), uncertainty concerning a system's microstate (classically, its phase space point) has come to be called 'information.' We wish to call attention to the crucial distinction between (i) epistemic and (2) ontological uncertainty or information, an issue that tends to be glossed over in discussions of thermodynamics and the Second Law. We note that epistemic uncertainty of a system's microstate, as assumed in classical statistical mechanics, arguably fails to rigorously uphold either the Second Law or Landauer's Principle (cf. Kastner 2017) and that ontological uncertainty is necessary for rigorous derivations of both.[1] This consideration leads to a restricted form of LP that does not rely on epistemic uncertainty as traditionally assumed. Essentially, the LP is indeed a form of the Second Law, but only insofar as both are independent of epistemic considerations and instead supervene on specific physics involving the ontology of the system, independently of human knowledge or lack thereof. In contrast, if LP is *defined* as applying to epistemic uncertainty, we concur with Earman and

---

[1] In particular, the fine-grained Gibbs entropy never increases. Thus if a system is genuinely in a particular microstate, and subject to unitary evolution, its entropy never increases, whether or not an outside observer knows what microstate it is in. The 2nd law is then only salvaged through 'coarse-graining,' an *ad hoc* device that conflicts with the given assumptions. It also implies (since $\Delta S = \Delta Q/T$) that the amount of heat and the temperature of a system depend on some outside observer's state of knowledge of the system (the 'degree of coarse graining'), which is absurd. One only obtains physically real entropy increase if coarse-graining accurately reflects the system's physical state, and this is only possible if the uncertainty is ontological.

Norton's refutations of the LP and provide an additional specific refutation based on an analysis of the actual physics involved in epistemic 'erasure' or 'reset' operations.

## 2. Background

Landauer provided some heuristic arguments in support of his principle, but ultimately appealed to the Second Law of Thermodynamics (Landauer 1961, 187). Earman and Norton (1999) have correctly pointed out that attempts to use Landauer's Principle (LP) to save the Second Law from violation by such devices as "Maxwell's Demon" are circular whenever they invoke the Second Law in support of LP. EN (1999) also argue that extant attempts to save the Second Law and/or LP without resort to the Second Law are subject to counterexamples and are not universal or conclusive. In addition, Norton (2013) argues that the information-theoretic approach to thwarting the Demon is similarly circular and fails to identify a fundamental and universal principle.

Meanwhile, Bennett (1982) claimed that the relevant irreversible process is that of erasure of memory devices rather than measurement, which he portrayed as copying. Bennett's formulation has become the Received View in the literature on this topic, and a consensus has developed that Bennett refuted Brillouin's identification of measurement as the source of the entropy cost attending the entropy-reducing sorting of the Demon (Brillouin, 1951). However, the copying model of measurement applies only to the classical limit and is thus inadequate at a fundamental level. If a system is prepared in a state $\psi(x)$ with spread $\Delta x$, measuring its position is an irreversible process that, by the uncertainty principle, expands the uncertainty associated with its momentum (and vice versa). It appears that the Received View has neglected to take this point into account, and it is the focus of the present work.

## 3. Epistemic Uncertainty Fails to Uphold Either 2nd Law or Landauer's Principle

Before dealing with further specifics, we first consider the important concern of Norton (2013) that one cannot simply assume that ``information entropy'' $I$ can be automatically identified with thermodynamic entropy $S$, a practice that has contributed to confusion and arguably to overstated claims about the findings of experiments (see, e.g., discussion of an experiment by Bérut *et al* (2012) in Norton (2013, 3.7). Here, we show that one cannot equate reset operations based only on epistemic uncertainty to phase space compressions (that are then putatively compensated by LP to save the Second Law). The tradition of doing so fails due to the neglect of crucial physics. For example, Ladyman and Robertson (2014, p. 2281) assume that free energy is expended in a reset instantiating LP based on the need to use a reset operation that is independent of the input state of the memory storage device. In fact, they regard this as the crucial meaning of a genuine "reset" operation that instantiates LP; they concede that LP fails if the reset is tailored to a specific memory state. However, the relevant physical content of the reset process is not exhausted by the reset operation itself but depends on the initial state of the system as well, as we show in the following analysis. Yet in any case, it turns out that negligible free energy is expended for either of the input states considered as determinate "facts of the matter" that are simply unknown to the experimenter. Thus, there is no "reset" in the real world that upholds the LP if the uncertainty about the memory state is taken as epistemic. In what follows, we explicate this situation.

First, we need to make a clear distinction between:
(a) the reset operation
(b) the interactions taking place between the reset device and the system.

To our knowledge, this distinction has not previously been made in the literature, largely because of the tradition of defining entropy in terms of an ignorance, rather than ontological, interpretation of probability. It is indeed true that our ignorance requires that we cannot tailor the reset to the initial state of the system, which leaves (a) independent of the initial state. However, the actual physics dictates that (b) depends on the initial state.

When these details are taken into account, we find that the reset does not in fact correspond to phase space compression, despite (a). The assumption that the actual physics depends on our ignorance of the state is the essential fallacy that leads to the illicit overgeneralization that equates a decrease of logical/epistemic information to a decrease of thermodynamic entropy.

Let us now consider the physics involved concerning the expenditure of free energy (if any) in a reset process. Suppose the memory device is a box of volume V containing a gas molecule that could be either on the left or right sides of the box; the input states are thus L or R.[2] The reset operation is the insertion of a piston from the right-hand side that imposes a constraint on the molecule such that only state L becomes available. This corresponds to (a), in that the reset operation must be independent of the initial state of the system for logical/computational reasons. First, we assume that there is no fact of the matter about the system's relevant microstate; specifically, it involves a quantum system whose L/R position is ontologically uncertain.

Recall that free energy $F$ for an isothermal process is defined by $dF = dE - TdS$, where $dS \geq \frac{dQ}{T}$.[3] The equality holds in the reversible limit, which we consider here for simplicity.[4] Consider a system in a heat bath at temperature T while being compressed isothermally, so that an amount of heat ΔQ leaves the system and is dumped into the heat bath. Under these conditions, using the First Law $\Delta E = \Delta W + \Delta Q$ and for work $\Delta W = p\Delta V$ exerted on the system, the free energy change of the system is:

$$\Delta F = p\,\Delta V + \Delta Q - T\Delta S = p\,\Delta V + \Delta Q - T\frac{\Delta Q}{T} = p\,\Delta V \quad (1)$$

The free energy expended in this process, $\Delta F_{exp}$ is thus defined by the work done on the system; specifically, $\Delta F_{exp} = p\,\Delta V$ (in the reversible limit). It is important to note that the above analysis assumes that the system's ontological uncertainty takes up the entire volume V, so that there is no fact of the matter about the system--it is not functioning as a memory state.

In contrast, suppose that the system *is* functioning as a memory such that it is actually in one of the two states L or R, and the uncertainty is merely epistemic. This is the situation that corresponds to the computational form of Landauer's Principle, advocated for example by Plenio and Vitelli (2001). They illustrate the situation with this figure:

---

[2] This is the same situation considered in extant discussions of LP; for example, Plenio and Vitelli (2001); Kycia and (2020); Feynman (1996). Thus, on pain of a double standard, the present discussion cannot be rejected based on any alleged inapplicability of entropy to single-particle systems.

[3] To be totally precise, in the case of irreversibility the dQ would be replaced by a δQ. We neglect this detail here.

[4] Norton (2017) correctly argues that perfect reversibility is an unphysical idealization.

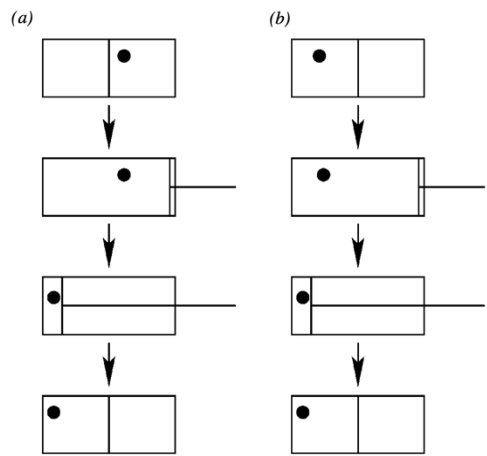

**Figure 3.** We erase the information of the position of the atom. First we extract the wall separating the two halves of the box. Then we use a piston to shift the atom to the left side of the box. After the procedure, the atom is on the left hand side of the box irrespective of its initial state. Note that the procedure has to work irrespective of whether the atom is initially on the right (*a*) or on the left side (*b*).

(Plenio and Vitelli, 2001,27; used with permission.)

They then say: "The physical result of the compression is a decrease in the thermodynamical entropy of the gas by *k ln 2*. The minimum work that we need to do on the box is *kT ln 2*, if the compression is isothermal and quasi-static." They then assume, by the 2nd Law, that this alleged decrease in the entropy of the gas must be compensated by a dissipation of at least *kT ln 2*, i.e., Landauer's Principle (computational/epistemic version).

However, they have not calculated the work done for the case in which the atom is truly in one particular microstate—i.e., either left-hand or right-hand side of the box, L or R—as a real memory system would be. Instead, they tacitly assume that the atom's uncertainty physically corresponds to the entire volume of the box once they remove the partition. (This seems to be a standard feature of extant 'derivations' of Landauer's principle in its computational form). However, were that the case, the gas would have undergone a free expansion that increased its entropy. Then, the depicted compression would not correspond to any net reduction of entropy of the gas of *kT ln 2*, but simply a return of the gas to its initial entropy state. In what follows, we correct this error by keeping in mind that the gas, functioning as a memory system, really is in one state or the other (L or R).

The actual physical interaction between the reset device and the system differs for the two different memory states—this corresponds to observation (b). Yet in any case there is no real physical compression going on in this situation; i.e., when the uncertainty ('information') is epistemic. Both these points can be demonstrated as follows. (We assume that the piston is applied a negligible time interval after partition removal.) If the molecule is L, then it is not in contact with the piston at all (in the piston's ready state). Thus, clearly $p = 0$, and the motion of the piston is merely *the placing of a constraint that does no work* $\Delta W = p\Delta V$ on the molecule, since $p \Delta V = 0$. The actual physical process occurring when the molecule is on the L is essentially no different from reducing the volume of an empty box, requiring zero thermodynamic work.

If the atom is in the state R, by assumption it is not taking up the entire volume *V* but only the right-hand region of volume $V/2$. Therefore, the action of the piston is not a compression of the system's physical uncertainty from $V$ to $V/2$; rather, $\Delta V = 0$ in this case. That is, the system is simply being pushed from one region of volume $V/2$ to a different region of volume $V/2$. Thus, while the specific physical interaction differs between the two memory states, still the relevant *thermodynamic* work $\Delta W = p\Delta V = 0$, even if there is a finite force F exerted against the atom in the case of state R. The latter serves merely to translate the position of the atom, while its associated phase space volume remains unchanged. While there will be some dissipation associated with this process relating to an amount of work to displace the system, $\Delta E = F\Delta x$, it does not amount to a non-negligible value for $p\Delta V$ and therefore does not uphold Landauer's principle (computational/epistemic form).

Thus, it is *not at all true* that "only if we do not know which side of the compartment the atom is in do we expend free energy," despite Feynman's having fallen into this trap as well (Feynman (1996, pp. 142–144). He succumbs to this fallacy because he defines entropy solely in terms of epistemic probability, which leads to the endemic but erroneous conflation of "ignorance about a system's state" with physical phase space volume. His statement overlooks the fact that free energy F is defined *not* by any observer's state of knowledge but by quantities of heat transferred and work performed by (or on) the system, and it is clear from the above analysis that zero work $p \Delta V$ is actually performed when the atom is in a definite memory state.

The conclusion is that one cannot equate 'erasure' of epistemic information in a reset operation to reduction of thermodynamic entropy, even if the reset is constrained to be the same operation applied regardless of the memory's initial state. This is simply not physically valid in view of the analysis above.

**4. Erasure of ontological uncertainty involves an entropy cost**

Nevertheless, the present work supports the relation of $I$ to $S$ under appropriate physical conditions. In teasing out the relation of $I$ to $S$, we must clearly distinguish between epistemic uncertainty (in which a system is in a determinate but unknown outcome-state) and ontological uncertainty (in which a system cannot be said to possess a property corresponding to an outcome-state). Let us denote the former, epistemic form of $I$ by $I_E$ and the latter, ontological form of $I$ by $I_O$. $I_E$, as Norton notes, needs to be distinguished from $S$: the fact that we do or don't happen to know which state a classical memory device is in does not affect its accessible phase space volume, and simply decreasing our ignorance cannot be equated to phase space compression on the part of the system itself.[6] This point is emphasized in the analysis of the previous section.

In contrast, $I_O$ is relevant to thermodynamic entropy due to the complementary relationship between incompatible observables in quantum systems and the fact that quantum states do not correspond to any determinate eigenvalue independently of measurement.[7] The concept of joint information applicable to the quantum case was studied by Hirschman (1957) and Leipnik (1959), who built on the pioneering work of Weyl (1928). These studies yielded a quantitative expression for the joint information for Fourier transform pairs such as position and momentum. Specifically, we have the relation:[8]

$$-\int_{-\infty}^{\infty}|\psi(x)|^2 \ln|\psi(x)|^2 \, dx - \int_{-\infty}^{\infty}|\varphi(p)|^2 \ln|\varphi(p)|^2 \, dp \geq \ln\left(\frac{he}{2}\right) \quad (2)$$

where the left-hand side is the joint information denoted $L$ in Leipnik's presentation. The associated dimensionless joint quantum information $I_O$ is defined as $I_O = L - \ln h$:

---

[6] This is why erasure of a classically-modeled memory device does not correspond to a true phase space compression, as Norton notes (2013, 3.6): "Erasure reduces logical space but not physical phase space." See also Norton (2005).

[7] Even under an assumption of hidden variables $\{\lambda\}$, $I_O$ applies to a quantum system in state $|\psi\rangle$, since a putative property-possession $\lambda$ does not define a state $|\psi\rangle$.

[8] The differential form can in principle yield negative values, but may be viewed as an idealization, since the notion of a position and momentum continuum can also be viewed as idealizations. In addition, there is no true position observable at the fully relativistic level, so that localization is likely limited to a Planck volume (cf. Schlatter and Kastner, 2023). Thus (2) may be viewed as an excellent approximation for the macroscopic level.

$$I_O = -\int_{-\infty}^{\infty}|\psi(x)|^2 \ln|\psi(x)|^2\, dx - \int_{-\infty}^{\infty}|\varphi(p)|^2 \ln|\varphi(p)|^2\, dp - \ln h \geq \ln\left(\frac{e}{2}\right) \quad (3)$$

Note that this quantity reflects the accessible position and momentum state spaces corresponding to an objective uncertainty or spread in values, since for general $\psi(x)$ there is no fact of the matter about the system's possession of any particular value of either observable. For $\psi(x)$ a Gaussian, (2) and (3) become equalities. In this case, it is clear that any reduction in the information associated with either observable must be compensated by an increase by the same amount in the complementary observable. I.e., a reduction in position-information $I_{O,x}$ forces an increase in momentum-information, $I_{O,p}$, and vice-versa. Thus, since thermal states correspond to Gaussian distributions, the relation (3) implies at least an information-cost for reducing the information associated with either observable of a complementary pair, for any quantum system in thermal equilibrium.

Now, how do these considerations connect to thermodynamic entropy? Recall that classical thermodynamic entropy can be defined in terms of Boltzmann's quantity H,

$$S = k_B I_{O,p} = -k_B \int_{-\infty}^{\infty}|\varphi(p)|^2 \ln|\varphi(p)|^2 dp, \quad (4)$$

where we make explicit that this definition is physically valid for ontological (not epistemic) uncertainty. In general, "information erasure" is identified with the position observable; for example, true compression of a gas to a smaller volume (in contrast to the "reset" operation based on epistemic uncertainty) reduces its ontological position-information $I_{O,x}$ and that is paid for by a momentum-based entropy cost as defined in (4). Arguably, (3) together with (4) are the genuine physical basis for the Second law. Insofar as Landauer's Principle can be related to the Second Law, this is the only sense in which Landauer's Principle is physically justified.

This point--that it is quantum uncertainty that yields the Second Law (and a restricted, ontological form of LP)--has been obscured in the literature because the states of quantum systems in thermal equilibrium are usually given only in terms of the energy (momentum) basis, and the position basis is neglected; thus, the constraint represented by (2) is generally not evident and not taken into account. Moreover, the typical state description is that of an ostensibly proper mixed state such as

$$\varrho = e^{-\beta \hat{H}}/Z, \quad (5)$$

which misrepresents the realistic situation in which a system is in a pure state of indeterminate energy (as in (2)), or in an improper mixed state (as in the case of entanglement with other quantum systems). In either case, it is not legitimate to view a subsystem as being in a determinate (pre-existing) but merely unknown energy state, so that measuring its energy is not a reversible copying procedure. Instead, measurement is, in general, irreversible.[9]

As a simple example of the relevance of (2), consider a single gas molecule in a large box of length $l$. This is a quantum system described to a good approximation by a pure state Gaussian $\psi_G(x, \sigma_x = l)$, such that its momentum space wave function is $\varphi_G\left(p, \sigma_p = \frac{h}{4\pi l}\right)$. Using the molecule to do work via a piston would require detecting which side of the box it occupies, an irreversible position measurement that reduces the x-uncertainty such that the resulting state is $\psi'_G(x, \sigma = l/2)$. By the uncertainty principle, there is an accompanying increase in the spread of the momentum space wave function such that $\varphi'_G\left(p, \sigma_p = \frac{h}{2\pi l}\right)$, and therefore the thermodynamic entropy (4) increases. This situation, rather than any considerations about classical/epistemic memory erasure (which, as noted earlier, does not actually affect physical entropy) or dissipation effects (to which there are arguably always exceptions, cf. Earman and Norton (1999)), is what enforces Landauer's Principle, but only insofar as it is applies to reduction of *ontological* uncertainty (ontological erasure).[10]

---

[9] Quantum measurement and its relation to entropy increase is discussed in detail in Kastner (2017). Of course, approaches assuming that quantum theory has only unitary evolution are subject to the same lack of true state space compression as the classical theory, which may be why the results offered herein have thus far been overlooked.

The increase in momentum-entropy corresponds to heating up the molecule, and indeed, in order to detect its position we must provide energy $\Delta Q$. This is the familiar observation that there is no detection without the involvement of photons, and that the more precisely we wish to measure the position, the higher the energy of the photon(s) required. Thus reduction in $I_{0,x}$ inevitably leads via (1) to an associated increase in entropy $\frac{\Delta Q}{T}$ that thwarts usage of the system to extract useful work.

For the isothermal case, in which the system's internal energy E remains constant in a heat bath at temperature T, we can derive the Landauer result for *ontological* erasure of one bit of position-information (halving of the volume) via the First Law of Thermodynamics as follows (using the ideal gas approximation):

$$dE = 0 = dW + dQ \Rightarrow dW = -dQ \qquad (6)$$

$$\int dW = \int p\, dV = \int_V^{V/2} \frac{kT}{V}\, dV = kT[\ln(V/2) - \ln V] = -kT\ln 2 \qquad (7)$$

$$\Rightarrow \Delta Q = kT\ln 2, \ \Delta S \geq k\ln 2 \qquad (8)$$

Again, however, this derivation is only valid for ontological uncertainty of the gas with respect to the volume, since no actual compression of the volume takes place in the case of epistemic uncertainty, as shown in Section 3.

The case of momentum measurement corresponds to that of the original Maxwell's Demon, who is conceived as sorting faster and slower molecules into each half of the box in order to violate the Second Law. It is traditionally assumed that the molecules have a well-defined but unknown momentum, but of course this is not the case at the quantum level; thus, there is never a "Maxwell's Demon" as Maxwell originally envisioned. Treatments that presuppose this classical picture must inevitably fall short of the relevant physics. Any real "Demon" must deal with the quantum nature of the gas molecules. In practice this means that any success the Demon has in measuring the momentum of a molecule must leave its position so indeterminate as to prevent him from getting it through the trap door. In other words, he now knows where to put it, but it has been so delocalized as to prevent physical sorting.

In more quantitative terms, the door width *d* must of the order of the position spread of the molecules to prevent leakage at temperature *T*, $d \sim \frac{h}{4\pi\sigma_p}$, where $\sigma_p \sim p_{rms} = \sqrt{3mkT}$. However, an accurate measurement of momentum involves a low-energy photon, $h\nu \ll kT$. This results in a position spread of the measured molecule of $\sigma_x \sim \frac{h}{4\pi\sqrt{mh\nu}} \gg d$. We thus note that exorcism of the Demon does not even require recourse to specific entropic considerations: the quantum uncertainty principle precludes the sorting required, so the question of "How is the entropy reduction from sorting compensated?"

---

[10] An example of the consequences of overlooking the uncertainty principle in measurement is illustrated by the discussion in Kycia and Niemczynowicz (2020), who assume that there is no entropy cost to a Demon's localization of the particle. This of course results in the paradox of the Demon's alleged ability to reduce entropy, which is then ostensibly cured by recourse to the computational/epistemic form of Landauer's Principle. But, as shown in the foregoing, that fails to do the job. In fact it is the uncertainty principle that prevents the existence of a real Demon.

never arises in the first place. The Demon cannot even begin to sort a quantum gas, and a real gas is a quantum gas.

Finally, it should be noted that the Second Law is defended from these sorts of challenges not by any argument about the specific pragmatics of thwarting the various mechanical manipulations involved in any particular scenario, but by the joint ontological information constraint (2), which is an inviolable property of any Fourier transform pair of which the quantum uncertainty principle is an example. This prevents, *in principle*, violation of the Second Law, and in particular provides a pre-emptive "no-go" for the Demon" that does not depend on illicit circular reference to the Second Law. The conclusion is that there can be no Maxwell's Demon in the real world.

5. Conclusion.

It has been shown that the information-theoretic form of Landauer's Principle, insofar as it equates epistemic uncertainty to thermodynamic entropy, fails due to neglect of relevant physics. Nevertheless, a restricted form of Landauer's Principle corresponding to ontological (as opposed to epistemic) information is implied for a thermal system by the joint entropy corresponding to the quantum uncertainty principle, independently of the Second Law. This remedies the current situation, in which Landauer's Principle has often been invoked as a surrogate for the Second Law when it is the Second Law itself that is being challenged. It also arguably provides the true physical basis for the Second Law, and reminds us why there can be no Maxwell's Demon.

In view of the findings herein, the question may naturally arise: why has the traditional epistemic treatment of probabilities in statistical mechanical measures of entropy, such the Boltzmann entropy $S_B = k \ln \Omega$, been empirically successful as quantifier of thermodynamic entropy? The reason is that $S_B$ applies to a system free to undergo rapid transitions among the microstates counted by W, such that no single microstate applies to the system over the time period to which $S_B$ applies; the system truly is in a macrostate and not one particular microstate over the relevant time period (recall the ergodic hypothesis). Moreover, the transitions are treated as inherently random (indeterministic) by $S_B$ under Boltzmann's *Stoßzahlansatz*, or "molecular chaos" assumption. The Stoßzahlansatz is technically inconsistent with the determinism of classical physics, but consistent with quantum probabilities taken as ontological. Thus, $S_B$ can be understood as applying to ontological uncertainty at a fundamental level, and this is why it is a valid measure of entropy despite the fact that it was introduced in a context in which it seemed that the only available form of probability was epistemic. We therefore find that the Stoßzahlansatz, which has been a longstanding point of contention concerning the legitimacy of Boltzmann's approach to entropy, turns out to be what saves his formula $S_B$ as a legitimate measure of entropy. It can now be seen as a feature of the physics that Boltzmann was able to discern as a requirement for a fully correct understanding of entropy $S = Q/T$. Heat $Q$, which is what allows transitions among microstates,[11] must be an indeterministically emerging quantity. This is consistent with the fact that $Q$ cannot truly be described by an exact differential (note 1).

In contrast, for the case of computation, in order to serve as a memory a system must remain in a stable *microstate* over the period of time applying to the calculation of information *I*. If *I* is then taken as a measure of the uncertainty over all possible microstates corresponding to some computational "macrostate", it is irreducibly epistemic, $I_E$, since the system *never really instantiates such a macrostate over the relevant time period*. This means that there is no heat involved (i.e., no changes of microstate), and that $I_E$ is not equivalent to *S*.

---

[11] Rovelli and Smerlak (2011) make explicit that heat corresponds to state transitions, in their definition of the concept of thermal time.

**Acknowledgments:** The authors gratefully acknowledge valuable correspondence with John Norton.